\documentstyle[11pt,newpasp,twoside,epsf]{article}
\def\edcomment#1{\iffalse\marginpar{\raggedright\sl#1\/}\else\relax\fi}
\reversemarginpar

\begin{document}
\title{Black hole masses of BL Lac objects}
 \author{J.K. Kotilainen}
\affil{Tuorla Observatory, University of Turku, V\"ais\"al\"antie 20, 
FIN--21500 Piikki\"o, Finland; e-mail: jarkot@astro.utu.fi}
\author{R. Falomo}
\affil{Osservatorio Astronomico di Padova, Vicolo dell'Osservatorio 5, 
35122 Padova, Italy}
\author{A. Treves}
\affil{Universit\`a dell'Insubria, via Valleggio 11, 22100 Como, Italy}

\begin{abstract}
The correlation between black hole mass $M_{BH}$ and the central stellar 
velocity dispersion $\sigma$ in nearby elliptical galaxies affords a 
novel way to determine $M_{BH}$ in active galaxies. 
We report on measurements of $\sigma$ from optical spectra of 7 BL Lac 
host galaxies. The range of $\sigma$ (160 -- 290 km s$^{-1}$) corresponds to 
estimated $M_{BH}$ of 5 $\times$ 10$^7$ -- 9 $\times$ 10$^8$ M$_\odot$. 
The average ratio of $M_{BH}$ to the host galaxy mass is 
1.4 $\times$ 10$^{-3}$, consistent with that estimated in other 
active and inactive galaxies. The range of $\sigma$ and $M_{BH}$ of the 
BL Lacs are similar to those obtained for low redshift radio galaxies, 
in good agreement with the predictions of the unified models for 
radio--loud active galaxies. 
\end{abstract}

\section{Introduction}

The mass of the central black hole (BH) is of paramount importance in 
theoretical models of AGN. In particular, the dependence of BH mass 
($M_{BH}$) on global host galaxy properties provides clues to the role of BHs 
in galaxy formation and evolution. Dynamical determination of $M_{BH}$ in AGN 
is difficult because of the bright nuclear emission. The main method that 
has proved to be successful for AGN is the time-consuming 
reverberation mapping of broad emission lines.
Only for a few quasars and Seyfert galaxies $M_{BH}$ is thus known 
(e.g. Kaspi et al. 2000; Wandel 2002). Reverberation mapping cannot obviously 
be employed for BL Lac objects because they lack prominent broad emission 
lines. The discovery of a correlation between $M_{BH}$ and the bulge 
luminosity in nearby early-type galaxies (e.g. Magorrian et al. 1998) offered 
a new tool for evaluating $M_{BH}$ (see Merritt \& Ferrarese 2001). 
This correlation has been applied so far 
for nearby quasars (McLure \& Dunlop 2001) and BL Lacs (Treves et al. 2002).

Recently, a tighter correlation was found relating $M_{BH}$ with the stellar 
velocity dispersion $\sigma$ of the bulge in nearby inactive galaxies 
(Gebhardt et al. 2000; Ferrarese \& Merritt 2000). This relationship clearly 
demonstrates a connection between BHs and bulges of galaxies and has spurred 
substantial effort in theoretical modelling (e.g. Silk \& Rees 1998; 
H\"ahnelt \& Kauffmann 2000). The relationship predicts more accurately 
$M_{BH}$, but requires the measurement of $\sigma$ in AGN host galaxies 
that is difficult, in particular for objects at high redshift and with 
luminous nuclei. On the other hand, BL Lacs have relatively fainter nuclei 
than quasars, and for low redshift BL Lacs this measurement can be secured 
with a single spectrum observable with a medium-sized telescope.

We have carried out medium resolution optical spectroscopy of the 
host galaxies of nearby (z $<$ 0.2) BL Lac objects to derive $\sigma$ and to 
determine $M_{BH}$. Here we present results based on a sample of 7 BL Lacs 
including 2 LBLs (3C 371 and PKS 2201+04) and 5 HBLs (see also Falomo, 
these proceedings; and full discussion in Falomo, Kotilainen \& Treves 2002),
For all of them, high quality images have been obtained either from 
the ground (Falomo \& Kotilainen 1999) or with HST (Urry et al. 2000; 
Falomo et al. 2000). From these images, the characterization of the 
host galaxies and the nuclear luminosity can be obtained. 
In Falomo, Kotilainen \& Treves (in prep.), we shall discuss the 
implications of the results of the full sample of 11 BL Lacs for the 
unified models of radio-loud AGN.

\section{Observations and data analysis}

The observations were obtained using the 2.5m Nordic Optical Telescope (NOT) 
equipped with ALFOSC. Spectra were secured using two grisms to cover the 
spectral ranges 4800 -- 5800 \AA~(setup A) and 5700 -- 8000 \AA ~(setup B) at 
0.54 \AA~pixel$^{-1}$ and 1.3 \AA ~pixel$^{-1}$ dispersion, respectively. 
This allows us to measure the absorption lines of H$\beta$ (4861 \AA), 
Mg I (5175 \AA), Ca E-band (5269 \AA), Na I (5892 \AA) and the 
TiO + CaI (6178 \AA), TiO + FeI (6266 \AA) and other absorption line blends 
from the host galaxies at a spectral resolution R $\sim$3000.

The chosen grisms combined with a 1\arcsec\ slit yield a spectral resolution 
for $\sigma$ measurement of $\sim$60 -- 80 km s$^{-1}$, which is 
adequate for the expected range of $\sigma$ in luminous ellipticals (e.g. 
Djorgovski \& Davis 1987; Bender, Burstein \& Faber 1992) such as the hosts 
of BL Lacs. In addition, we acquired spectra of bright stars of type G8-III 
to K1-III, that exhibit low rotational velocity 
(V$\times$$\sin{(i)}<$ 20 km$s^{-1}$) to be used as templates of zero 
velocity dispersion. Furthermore, spectra of the well studied nearby 
elliptical galaxy NGC 5831 were secured to provide a test of the adopted 
procedure to derive $\sigma$.

During the observations, seeing ranged between 1\arcsec\ and 1.5\arcsec . \ 
The slit was centered on-target or positioned 1\arcsec\ away from the nucleus 
and the 1D spectrum was extracted from an aperture of 3\arcsec\ - 5\arcsec\ 
diameter, in all cases within the effective radius of the host galaxy. In one 
case (Mrk 180), spectra with the slit both on-target and off-centered by 
1\arcsec\ were taken but no significant difference was found in the shape of 
the spectral features. 

The stellar velocity dispersion $\sigma$ was determined using the 
Fourier Quotient method (e.g. Sargent et al. 1977). The spectra were 
normalized by subtracting the continuum, converted to a logarithmic scale 
and multiplied by a cosine bell function that apodizes 10\% of the pixels 
at each end of the spectrum. All regions affected by emission lines 
(see Fig. 1) have been masked in the analysis. Finally, the Fourier Transform 
of the galaxy spectra were divided by the Fourier Transform of template stars 
and $\sigma$ was computed from a $\chi^{2}$ fit with a Gaussian broadening 
function (see e.g. Bertola et al. 1984; Kuijken \& Merrifield 1993). 
The $rms$ scatter of the $\sigma$ results using different template stars was 
typically $\sim$10 km$s^{-1}$ and can be considered as the minimum 
uncertainty of the measurement. 

For three objects we have spectra in both spectral ranges. The resulting 
values of $\sigma$ are in good agreement, ensuring sufficient homogeneity of 
data taken with different grisms and/or resolution. Note, however, that there 
is a tendency for the red grism (lower resolution) data to result in slightly 
larger value of $\sigma$. For the nearby elliptical NGC 5831 we obtained 
$\sigma$ = 167$\pm$5 and 185$\pm$10 km s$^{-1}$ for the setup A and B, 
respectively, in good agreement with previous measurements 
($<\sigma>$ = 168 km s$^{-1}$; Prugniel et al. 1998). In Fig. 1 we show the 
spectrum of the BL Lac object PKS 2201+04, observed in both spectral ranges.

\begin{figure}[ht]
\plotone{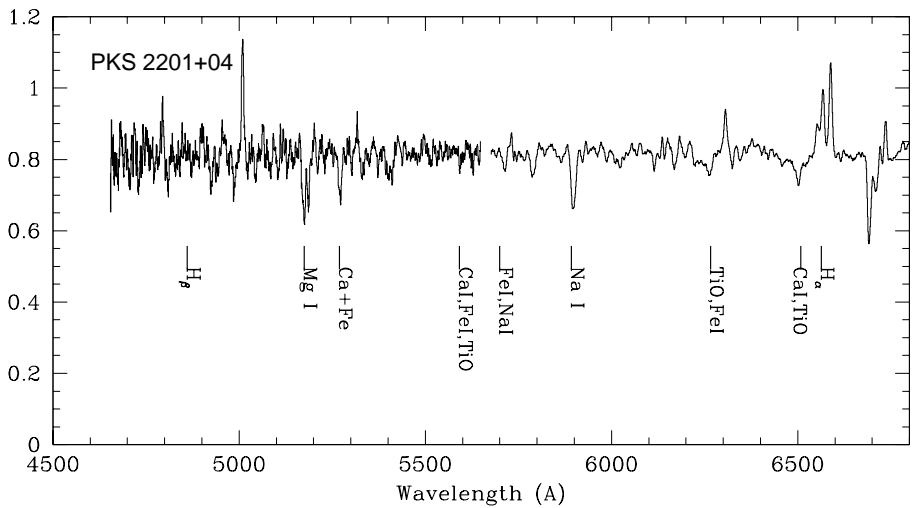}
\caption{Optical spectrum of the BL Lac object PKS 2201+04 (z = 0.027), 
normalized to the continuum and plotted in the rest frame.}
\end{figure}

Since early-type galaxies exhibit gradients in velocity dispersion (e.g. 
Fisher, Illingworth \& Franx 1995), the measured value of $\sigma$ depends 
somewhat on the distance of the galaxies and the aperture size 
(see Falomo et al. 2002 for details). The individual measurements of 
corrected $\sigma$ are given in Table 1. 

\begin{table}
\caption{Velocity dispersion and BH masses of BL Lacs.}
\begin{tabular}{llllll}
\tableline
Object & z & $\sigma_c$ & log($M_{BH}$)$_\sigma$ & log($M_{BH}$)$_{bulge}$ & log($M_{(host)}$) \\
& & km s$^{-1}$ & [M$_{\odot}$] & [M$_{\odot}$] & [M$_{\odot}$] \\ 
Mrk 421     & 0.031 & 236$\pm$10& 8.50$\pm$0.18 & 8.65 & 11.20 \\
Mrk 180     & 0.045 & 244$\pm$10& 8.57$\pm$0.19 & 8.50 & 11.45 \\
Mrk 501     & 0.034 & 291$\pm$13& 8.93$\pm$0.21 & 9.00 & 11.59 \\
I Zw 187    & 0.055 & 253$\pm$15& 8.65$\pm$0.18 & 8.20 & 11.39 \\
3C 371      & 0.051 & 284$\pm$18& 8.88$\pm$0.20 & 8.90 & 11.32 \\
1ES 1959+65 & 0.048 & 195$\pm$15& 8.12$\pm$0.13 & 8.30 & 11.27 \\
PKS 2201+04 & 0.027 & 160$\pm$7 & 7.72$\pm$0.13 & 8.27 & 11.00 \\
\end{tabular}
\end{table}

Barth, Ho \& Sargent (2002) have recently reported a similar systematic study 
of $\sigma$ in 11 BL Lac objects, six of which are common with our sample. 
There are some discrepancies for individual BL Lacs, most notably for Mrk 501 
and I Zw 187, for which they derive $\sigma$ = 372$\pm$18 and 
171$\pm$12 km s$^{-1}$, respectively. These differences are most likely 
related to the method of deriving $\sigma$ (Fourier Quotient vs. 
direct fitting) and the different wavelength range used. The average 
values of $\sigma$, however, are in good agreement in the two samples.

\section{Results and discussion}

We have adopted the relationship between $M_{BH}$ and $\sigma_c$ found for 
nearby early-type galaxies based on optical spectroscopy 
(Merritt \& Ferrarese 2001): 
$M_{BH}$ = 1.48$\pm$0.24 $\times$ 10$^8$ ($\sigma$/200 km s$^{-1}$)$^{4.65\pm0.48}$ [M$_{\sun}$]. 
We assume that this relationship is also valid for BL Lacs. This is 
consistent with imaging studies of BL Lacs (Falomo \& Kotilainen 1999; 
Urry et al. 2000; Falomo et al 2000), indicating that all BL Lacs are hosted 
by giant ellipticals. 
The derived values of $M_{BH}$ are reported in 
Table 1, where the errors are the composition in quadrature of uncertainties 
in $\sigma$ and in the Merritt \& Ferrarese (2001) relationship. 
Using instead the Gebhardt et al. (2000) relationship tends to yield 
slightly lower values of 
$M_{BH}$ but does not substantially modify our main conclusions. The values 
of $M_{BH}$ (Table 1) span a factor $\sim$20 from 5 x 10$^7$ M$_\odot$ for 
PKS 2201+04 to 9 x 10$^8$ M$_\odot$ for Mrk 501, with a median value of 
4 x 10$^8$ M$_\odot$. These values are in good agreement with those derived 
by Barth et al. (2002).

As mentioned above, $M_{BH}$ is also correlated with the bulge luminosity 
of the host galaxy. $M_{BH}$ was calculated following the relationship by 
McLure \& Dunlop (2002): 
log $M_{BH}$ = --0.50$\pm$0.05 M$_R$ -- 2.91$\pm$1.23 [M$_{\odot}$].
The corresponding values of $M_{BH}$ are given in Table 1. For most sources 
the difference of $M_{BH}$ derived with the two methods is within the 
estimated uncertainty. The average values of $M_{BH}$ derived, respectively, 
from $\sigma$ and the host luminosity are: 
$<$log$ M_{BH}>_{\sigma}$ = 8.62 $\pm$ 0.23 and 
$<$log$ M_{BH}>_{host}$ = 8.66 $\pm$ 0.25.

The measurements of $\sigma$ combined with the effective radii of the host 
galaxies can be used to estimate the mass of the hosts through the 
relationship (Bender et al. 1992): $M_{host}$=5$\sigma^{2}r_e$/G. 
This dynamical mass (Table 1) is in the range of 
1 -- 4 x 10$^{11}$ M$_\odot$. The ratio between $M_{BH}$ and $M_{host}$ is in 
the range of 0.5 -- 3.6 $\times$ 10$^{-3}$, with average 
$<M_{BH}$/$M_{host}>$ = 1.4 $\times$ 10$^{-3}$. This is in good agreement 
with values derived for both AGN and inactive galaxies 
($<M_{BH}$/$M_{host}>$ = 1.2 $\times$ 10$^{-3}$ ; McLure \& Dunlop 2001; 
Merritt \& Ferrarese 2001).

According to the unified model of radio-loud AGN (e.g. 
Urry \& Padovani 1995), BL Lacs are drawn from the population of FR I 
radio galaxies viewed along the jet axis. It is therefore interesting to 
compare orientation-independent properties of BL Lacs and radio galaxies. 
The largest comparison sample with available measurements of $\sigma$, 
effective radii and surface brightness is that of 73 low redshift 
radio galaxies by Bettoni et al. (2001, 2002). In Fig. 2, we compare the 
distribution of the BH masses of the BL Lacs and the radio galaxies from 
Bettoni et al.

\begin{figure}[ht]
\plotone{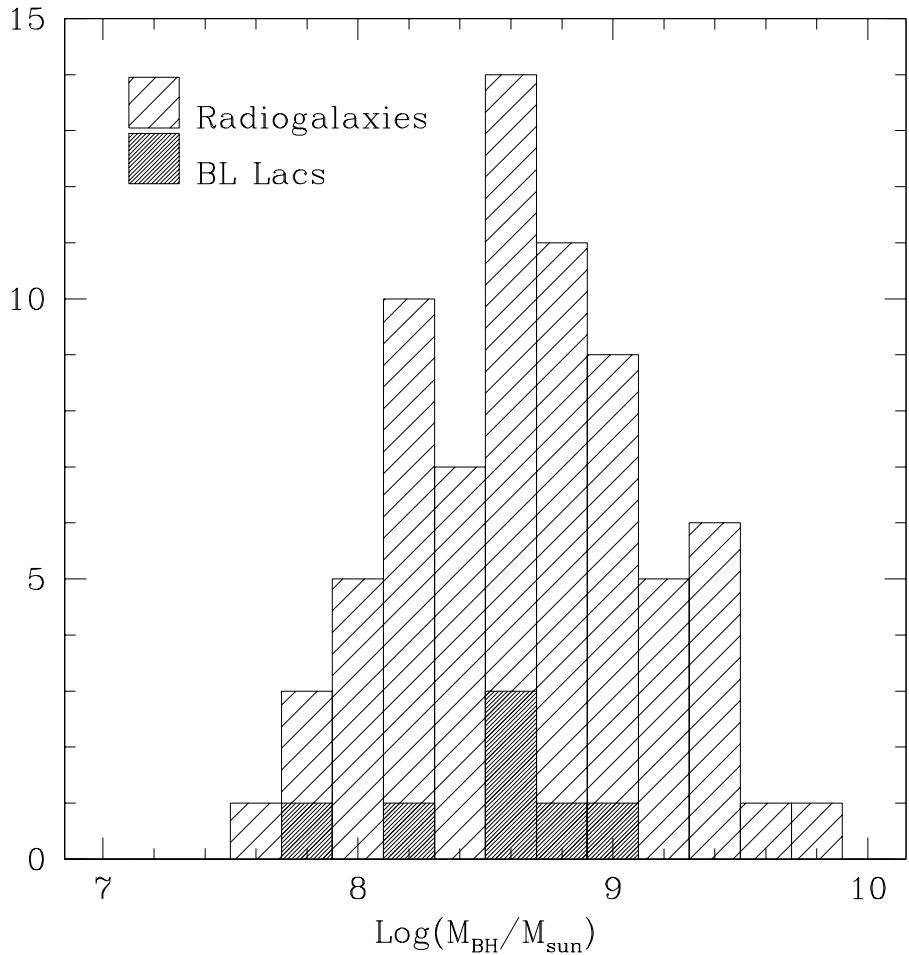}
\caption{The distribution of the $M_{BH}$ of the BL Lac objects compared with 
that of low redshift radio galaxies (from Bettoni et al. 2002).}
\end{figure}

The close similarity of the distributions in these samples implies that the 
distributions of $M_{BH}$ in BL Lacs and radio galaxies are 
indistinguishable, consistent with both types of AGN belonging to the same 
parent population, observed from different orientation angles (see also 
Barth et al. 2002). Additional support to this view is given by the 
similarity of BL Lacs and radio galaxies on the fundamental plane (see e.g. 
Falomo, these proceedings; Barth et al. 2002). However, due to the 
small number statistics of the BL Lacs, and the incompleteness of 
the samples, this result can not yet be considered conclusive. 
Furthermore, there is a clear dearth of BL Lacs with the largest 
$M_{BH}$ ($>$ 10$^9$ M$_\odot$), consistent with BL Lac hosts being biased 
toward less massive and less luminous host galaxies than radio galaxies 
(e.g. Urry et al. 2000).


\begin{references}
\reference{}Barth,A., Ho,L.C., Sargent,W.L.W., 2002, ApJ, in press (astro-ph/0209562)
\reference{}Bender,R., Burstein,D., Faber,S.M., 1992, ApJ 399, 462
\reference{}Bertola,F., Bettoni,D., Rusconi,L., Sedmak,G., 1984, AJ 89, 356
\reference{}Bettoni,D., Falomo,R., Fasano,G., et al., 2001, A\&A 380, 471
\reference{}Bettoni,D., Falomo,R., Fasano,G., Govoni,F., 2002, A\&A, in press
\reference{}Djorgovski,S., Davis,M., 1987, ApJ 313, 59
\reference{}Falomo,R., Kotilainen,J., 1999, A\&A 352, 85
\reference{}Falomo,R., Scarpa,R., Treves,A., Urry,C.M., 2000, ApJ 542, 731
\reference{}Falomo,R., Kotilainen,J.K., Treves,A., 2002, ApJ 569, L35
\reference{}Ferrarese,L., Merritt,D., 2000, ApJ 539, L9
\reference{}Fisher,D., Illingworth,G., Franx,M., 1995, ApJ 438, 539
\reference{}Gebhardt,K., Bender,R., Bower,G., et al., 2000, ApJ 539, L13
\reference{}H\"ahnelt,M.G., Kauffmann,G., 2000, MNRAS 318, L35 
\reference{}Kaspi,S., Smith,P.S., Netzer,H., et al., 2000, ApJ 533, 631 
\reference{}Kuijken,K., Merrifield,M.R., 1993, MNRAS 264, 712
\reference{}Magorrian,J., Tremaine,S., Richstone,D., et al., 1998, AJ 115, 2285
\reference{}McLure,R., Dunlop,J., 2001, MNRAS 327, 199
\reference{}McLure,R., Dunlop,J., 2002, MNRAS 331, 795
\reference{}Merritt,D., Ferrarese,L., 2001, MNRAS 320, L30
\reference{}Prugniel,Ph., Zasov,A., Busarello,G., Simien,F., 1998, A\&AS 127, 117
\reference{}Sargent,W.L.W., Schechter,P.L., Boksenberg,A., Shortridge,K., 1977, ApJ 212, 362
\reference{}Silk,J., Rees,M.J., 1998, A\&A 331, L1
\reference{}Treves,A., Carangelo,N., Falomo,R., et al., 2002, Issues in Unification of AGN, eds. R.Maiolino et al., p. 303 (ASP)
\reference{}Urry,C.M., Padovani,P., 1995, PASP 107, 803
\reference{}Urry,C.M., Scarpa,R., O'Dowd,M., et al., 2000, ApJ 532, 816
\reference{}Wandel,A., 2002, ApJ, 565, 762
\end{references}
\end{document}